\documentclass[prl,twocolumn,showpacs,preprintnumbers,amsmath,amssymb]{revtex
4}

\usepackage{graphicx}
\usepackage{dcolumn}
\usepackage{bm}

\def\E{{\cal E}}
\begin{document}

\preprint{{\it Phys. Rev. Lett.}}

\title{Cosmic Microwave Radiation Anisotropies in Brane Worlds}

\author{Kazuya Koyama}
 
\affiliation{
Department of Physics, University of Tokyo 
7-3-1 Hongo, Bunkyo, Tokyo 113-0033, Japan
}

\date{\today}

\begin{abstract}
We propose a new formulation to calculate the Cosmic Microwave Background (CMB) 
spectrum in the Randall Sundrum two-branes model based on recent progresses 
in solving the bulk geometry using a low energy approximation. 
The evolution of the anisotropic stress imprinted on the brane by 
the 5D Weyl tensor is calculated. An impact of the dark radiation 
perturbation on CMB spectrum is investigated in a simple model
assuming an initially scale-invariant adiabatic perturbations. 
The dark radiation perturbation induces isocurvature perturbations, 
but the resultant spectrum can be quite different from the prediction of simple 
mixtures of adiabatic and isocurvature perturbations due to Weyl anisotropic stress.

\end{abstract}


\maketitle
\hspace{1cm}\\
{\it 1.Introduction}\\
By suggestions from string theory/M-theory,
much attention is paid to brane world ideas
where we are living on a 3-brane in higher-dimensional
spacetime. 
In these brane world models, only gravity propagates
in a higher-dimensional "bulk" spacetime while standard
model fields are confined to the brane. The simplest
realization of this idea was given by Randall and Sundrum \cite{RS}.
The 5D bulk has a negative cosmological constant 
$\Lambda_5=-6/l^2$ where $l$ is the curvature radius in the 
bulk and the brane has a tension $\lambda=6/\kappa_5^2 l$
where $\kappa_5^2=8 \pi G_5$ and $G_5$ is the 5D Planck 
constant.

In this model, predictions of 4D 
general relativity are modified due to the 
influence of the gravitational fields in the bulk. 
These modifications need to be consistent with 
cosmological observations. 
Among them, the observations on Cosmic Microwave Background 
(CMB) anisotropies are now dramatically
improving. The WMAP experiment has already provided precise  
measurements of CMB anisotropies \cite{WMAP}. 
It is then necessary to calculate the CMB spectrum in brane worlds
in order to check the consistency of the model.

These works were initiated soon after the paper of Randall Sundrum 
\cite{perturbations}.
Although a large number of papers have tried to give a prediction
of CMB anisotropy, there is still no quantitative
estimation. This is due to the difficulties in solving full
5D perturbations \cite{review}.
In view of observational improvements, it is eagerly
desired to develop a formulation which enables us to provide quantitative
predictions. In this Letter, we propose such a formulation 
based on recent progresses in solving bulk geometry using a low energy 
approximation \cite{low,Koyama,low2,ShiromizuKoyama}. 

We consider a simple Randall-Sundrum type two-branes 
model where a low energy approximation is applicable. The stabilization 
mechanism is not introduced and the physical brane is assumed to be the 
positive tension brane. This simple setup enables us to understand 
clearly how the bulk gravitational fields affect the evolution of 
perturbations on the brane. 

 \hspace{1cm}\\ 
{\it 2. View from the brane}\\ 
The effective 4D Einstein equation on the brane is given by\cite{SMS}:
\begin{equation}
G_{\mu \nu}=\kappa_4^2 T_{\mu \nu} +\kappa_5^2 \Pi_{\mu \nu}
-\E_{\mu \nu},
\label{4deinstein}
\end{equation}
where $\kappa_4^2 = 8 \pi G_4=\kappa_5^2/l$ and $\Pi_{\mu \nu}$
is the quadratic function of the energy momentum tensor, which
can be neglected at low energies $T^{\mu}_{\nu}/\lambda \ll 1$. 
The energy-momentum tensor satisfies the conservation's law 
$\nabla^{\mu} T_{\mu \nu}=0$ where
$\nabla^{\mu}$ is the covariant derivative on the brane.
$\E_{\mu \nu}$ is the 
electric part of the 5D Weyl tensor and carries the 
information in the bulk. From the 4D Bianci identities and  
the symmetry of Weyl tensor, $\E_{\mu \nu}$ should satisfy 
\begin{equation}
\nabla^{\mu} \E_{\mu \nu}=0,
\quad \E^{\mu}_{\mu}=0,
\label{4dbianchi}
\end{equation}
on the brane for $T_{\mu \nu}/\lambda \ll 1$. An important point is that
$\E_{\mu \nu}$ cannot be determined only from equations on the brane 
Eq.(\ref{4dbianchi}). One needs 5D equations for $\E_{\mu \nu}$ in the bulk
and this is the source of great complexity of the problem.
To observe this fact,  it is convenient to parameterize $\E_{\mu \nu}$ 
as an effective energy-momentum tensor;
\begin{equation}
-\E^\mu_\nu
 = \kappa_4^2
\left[
\begin{array}{ccc}
-(\rho_\E+\delta\rho_\E Y) & & a V_{\E} Y_i  \\&&\\
-a^{-1} V_{\E} Y^j & & (P_\E+\delta P_\E)\delta^i_j
 + \delta \pi_{\E} Y_{ij}
\end{array}
\right] \, \nonumber,
\label{Emunu}
\end{equation}
where $Y(k,x) \propto e^{i k x}$ is the normalized scalar harmonics and 
the vector $Y_i$ and traceless tensor $Y_{ij}$ are constructed 
from $Y$ as $Y_i=- k^{-1} Y_{,i} , \quad Y_{ij}= k^{-2} Y_{,ij}
+\delta_{ij} Y/3$. 

In the background universe, Eq.(\ref{4dbianchi}) gives
$\dot{\rho}_{\E}+4 H \rho_{\E}=0$ and $P_{\E}=\rho_{\E}/3$,
where $H=\dot{a}/a$ and $a$ is the scale factor.
The solution for Weyl energy density is then given by
$\rho_{\E} =C a^{-4}$ where $C$ is the integration constant.
Thus the contribution from $\E_{\mu \nu}$ appears as dark radiation.
It has been shown that $C$ is related to the mass of the 
AdS-Schwarzshild BH in the bulk. In the following, 
we will assume $\rho_{\E}=0$ in the background spacetime. 

Let us consider the perturbations around the above background 
spacetime. The linear scalar metric is taken as  
\begin{equation}
ds^2=-(1+2 \Psi(t) Y) dt^2 +a(t)^2  (1+2 \Phi(t) Y) 
\delta_{ij} dx^i dx^j.
\nonumber
\end{equation}
Eq.(\ref{4dbianchi}) give
\begin{eqnarray}
\dot{\delta \rho}_{\E} +4 H \delta \rho_{\E} &=& -a^{-1} k V_{\E},
\quad \delta P_{\E}=\frac{1}{3} \delta \rho_{\E}, \nonumber\\
 \dot{V}_{\E} +4 H V_{\E} &=& \frac{k a^{-1}}{3}
 (\delta \rho_{\E} - 2 \delta \pi_{\E}),
\label{eqP}
\end{eqnarray}  
for $\rho/\lambda \ll 1$. At large scales $k a^{-1}/H \to 0$, 
we can neglect the terms proportional to $k$. 
The solution for $\delta \rho_{\E}$ is given by
$\delta \rho_{\E} =\delta C a^{-4}$ where $\delta C$ is the 
integration constant. The amplitude $\delta C$ is related to the 
perturbatively small AdS-Schwarzshild mass in the bulk. 
A problem is that it is impossible to determine anisotropic stress 
$\delta \pi_{\E}$ because it is dropped from Eq.(\ref{eqP}) for $k a^{-1}/H \to 0$.
In \cite{large}, it was clearly shown that this uncertainty of 
$\delta \pi_{\E}$ prevents us from predicting CMB anisotropy. 
The temperature anisotropy caused by (direct) Sachs-Wolfe effect
is given by the formula \cite{large}
\begin{equation}
\frac{\Delta T}{T} = \Theta_0+\Psi=\zeta + \Psi - \Phi,
\label{SW}
\end{equation}
where $\Theta_0$ is the temperature anisotropy of radiation and 
we used Eq.(\ref{4deinstein}) in the second equality. 
The evolution of $\zeta=\Phi+\delta \rho/3(\rho+P)$
is determined only by the conservation 
of the matter energy-momentum tensor. We will assume the matter 
perturbation is adiabatic $\delta P=c_s^2 \delta \rho$ where 
$\delta \rho$ is the density perturbation, $\delta P$ is the pressure 
perturbation and $c_s$ is the sound velocity. The continuity equation 
for matter perturbation implies $\zeta=\zeta_{\ast}=const.$
In addition to $\zeta$, the solutions for metric perturbations 
$\Phi$ and $\Psi$ are needed. From the 
effective 4D equation (\ref{4deinstein}), we can get the equations
to determine $\Phi$ and $\Psi$ as 
\begin{equation}
\zeta_{tot} = \Phi - \frac{H}{\dot{H}} \left( 
\frac{\dot{\Phi}}{H}-\Psi \right),
\Phi+\Psi = -\kappa_4^2 k^{-2} a^2 \delta \pi_{\E},
\label{zetatot}
\end{equation}
where $\zeta_{tot}$ is the curvature perturbation on hypersurface of 
uniform total energy density;
\begin{equation}
\zeta_{tot}=\zeta + \frac{\delta \rho_{\E}}{3(\rho+P)}=
\zeta_{\ast}+ \frac{\delta C a^{-4}}{3 (\rho+p)}.
\label{curvtot}
\end{equation}
Note that the solution for $\zeta_{tot}$ was obtained only by 
the law of conservation. However, the latter equation 
in Eq.(\ref{zetatot}) contains undetermined $\delta \pi_{\E}$. 
Indeed, Eq.(\ref{SW}) is written at the decoupling time as 
\begin{eqnarray}
\frac{\Delta T}{T} &=& -\frac{1}{5} \zeta_{\ast}
-\frac{2}{3} \left(\frac{\rho_r}{\rho} \right) \delta C_{\ast}
\nonumber\\
&-&\kappa_4^2 k^{-2} a^2 \delta \pi_{\E}+2 \kappa_4^2 k^{-2}
a^{-5/2}\int a^{7/2} \delta \pi_{\E} da,
\label{SWaniso}
\end{eqnarray}
where $\delta C_{\ast} = \delta \rho_{\E}/\rho_r$ and 
$\rho_r$ is the radiation energy density.
The second term comes from the entropy perturbation
contribution due to Weyl energy density. Unless the behavior
of $\delta \pi_{\E}$ is known, we cannot say anything 
about the effect of this dark radiation. 
It is possible to assume some ansatz for $\delta \pi_{\E}$,
say $\delta \pi_{\E} \propto \delta \rho_{\E}$ \cite{BM}. 
However, because $\delta \pi_{\E}$ is induced by 5D gravitational
perturbations, it is essential to solve the evolution equations for 
$\E_{\mu \nu}$ in the bulk. 

\hspace{1cm}\\
{\it 3. Solution for $\E_{\mu \nu}$ in the bulk}\\
The evolution equations for $\E_{\mu \nu}$ obtained from 
5D Bianchi identity are consisted of discouragingly 
complicated partial differential equations.
Fortunately, the curvature radius on the brane defined
by $L^2 \sim  1/\nabla_{\mu}^2 \sim 1/\kappa_4^2 T^{\mu}_{\nu}$ 
at the decoupling time is significantly longer
than the curvature radius in the bulk $l$ because the
experiments on Newton's force law already impose
the constraint on $l$ as $l < 1$ mm. Thus the system has
a natural small parameter $\epsilon=(l/L)^2 \sim T^{\mu}_{\nu}/\lambda$ 
and it is possible to solve the equations by the perturbation in
terms of $\epsilon$. 
If we consider a second brane with negative tension 
$\lambda_c= -6/\kappa_5^2 l$ at the physical distance
$d_0(x) l$ from our brane, 
$\E_{\mu \nu}$ can be determined as \cite{low2,ShiromizuKoyama}
\begin{eqnarray}
\E^{\mu}_{\nu} &=& \frac{2}{e^{2d_0}-1} \left[ 
-\frac{\kappa_4^2}{2}( T^{\mu}_{\nu}+e^{-2 d_0}T^{\mu}_{c\: \nu})
- \nabla^{\mu}\nabla_{\nu} d_0 \right. \nonumber\\
&+& \left.\delta^{\mu}_{\nu} \nabla^2 d_0 
-\left( \nabla^{\mu}d_0 \nabla_{\nu} d_0 +
\frac{1}{2} \delta^{\mu}_{\nu} (\nabla d_0)^2 \right)
\right],
\label{solE}
\end{eqnarray}
where $T^{\mu}_{c \:\nu}$ is the energy-momentum tensor 
on the second brane. Because the bulk spacetime shrinks 
exponentially due to the negative cosmological constant
in the bulk, the curvature radius on the second brane 
$L_c$ is shorter for larger $d_0$ as $L_c=L e^{-d_0}$. 
Thus in order to ensure that the low energy approximation can be applied 
on the second brane $(l/L_c)^2 < 1$, the radion $d_0$ 
should satisfy $e^{-2 d_0} > (l/L)^2 \ll 1 $. 

The evolution of $d_0$ can be calculated 
from the traceless condition $\E^{\mu}_{\mu}=0$ in Eq.(\ref{4dbianchi});
\begin{equation}
\nabla^2 d_0 -(\nabla d_0)^2 = \frac{\kappa_4^2}{6}
\left( T+e^{-2 d_0}T_c \right).
\label{radion}
\end{equation}
From Eqs.(\ref{solE}) and (\ref{radion}) the 
behavior of $\E^{\mu}_{\nu}$ is completely determined. 
Substituting the solution for $\E^{\mu}_{\nu}$
into Eq. (\ref{4deinstein}), the effective theory on the brane becomes 
quasi-scalar-tensor theory \cite{ST}. 

\hspace{1cm}\\
{\it 4. View from the bulk}\\
The solution for $\E_{\mu \nu}$ (Eq.(\ref{solE})) should be consistent with 
Eq.(\ref{4dbianchi}). In the background spacetime, $\rho_{\E}$ 
is written in terms of $d_0$ and energy densities on both 
branes;
\begin{equation}
\kappa^2_4 \rho_{\E}= -\frac{1}{e^{2 d_0}-1} \left(6 H \dot{d}_0 -
3 \dot{d}_0^2- \kappa_4^2(\rho+ e^{-2 d_0}\rho_c)
\right) \nonumber.
\label{rhoEB}
\end{equation}
The evolution equation for $d_0$ can be integrated once with 
the help of the energy-momentum conservation on each brane. 
We get $\rho_{\E}=C a^{-4}$ where $C$ is the integration constant. 
In this case, the integration constant is interpreted as the initial 
condition for the radion $d_0(t)$. 
Now we turn to the perturbations. We denote the perturbation of the 
radion as $d_0(t,x)=d_0(t)+N(t)Y(k,x)$. $\delta \rho_{\E}$ can be evaluated 
in the same way as the background spacetime. We get 
$\delta \rho_{\E} =\delta C a^{-4}$ at large scales
where $\delta C$ is the integration constant associated
with $N$. An advantage of our approach is that an equation for 
Weyl anisotropic stress $\delta \pi_{\E}$ can be derived as
\begin{equation}
\kappa_4^2 a^2 k^{-2} \delta \pi_{\epsilon}= 
\frac{2}{e^{2 d_0}-1} N.
\label{SolpiE}
\end{equation}
Hence the behavior of $\delta \pi_{\E}$ is determined by 
the radion perturbation $N$. Because the radion
perturbation $N$ is coupled to the metric perturbations $\Phi$
and $\Psi$, we should solve them at the same time. The
behavior of the metric perturbations is determined by
imposing the equation of state on the matter perturbations:
$\delta P=c_s^2 \delta \rho$ and $\delta P_c=c_{s c}^2 \delta \rho_c$.
We then have three equations, $\delta P=c_s^2 \delta \rho$, 
$\delta P_c=c_{s c}^2 \delta \rho_c$ and 
$\Phi+\Psi=-\kappa_4^2 k^{-2} a^2 \delta \pi_{\E}$
for three unknown functions $N ,\Psi$ and $\Phi$. 
It is convenient to introduce a new set
of variables $\varphi, \varphi_c$ and $E$ to solve these equations 
\cite{Koyama}. We take 
\begin{eqnarray}
\Phi &=& -\varphi-k^{-2} a^2 H \dot{E}+\frac{1}{3}E,\:\:
N = \varphi_c-\varphi-k^{-2}a^2 \dot{d}_0 \dot{E},\nonumber\\
\Psi &=& - \varphi -k^{-2} a^2 (\ddot{E}+2 H \dot{E}). 
\label{parameter}
\end{eqnarray}
The equation $\delta P=c_s^2 \delta \rho$ gives the equation for $\varphi$;
\begin{eqnarray}
&&\!\!\!\!\!\!\!\!\!\!\!\!\!\!\!\!\!\!\!
\ddot{\varphi} +(2 +3c_s^2)H \dot{\varphi}-(3H^2+2 \dot{H}
+3 c_s^2 H^2) \varphi 
- \frac{1}{2}\left(\frac{1}{3}-c_s^2 \right) \kappa_4^2 
\delta \rho_{\E} \nonumber\\
&=& -k^2 a^{-2} \left[ \left(c_s^2+\frac{2}{3} \right) \varphi
- \frac{1}{3} \left(c_s^2+\frac{1}{3} \right) E
\right].
\label{eqphi0}
\end{eqnarray} 
At large scales $k a^{-1}/H \to 0$, 
the equation for $\varphi$ is decoupled from $E$ and it can be integrated once 
to give the first order differential equation for $\varphi$;
\begin{equation}
-\left[ \varphi-\frac{H^2}{\dot{H}} \left(\frac{\dot{\varphi}}{H}-
\varphi \right) \right]=\zeta_{\ast}+\frac{\delta C a^{-4}}{3 (\rho+P)},
\label{eqphi}
\end{equation}
where $\zeta_{\ast}$ is the integration constant.
We should note that the left-hand side of Eq.(\ref{eqphi}) is nothing but 
the solution for $\zeta_{tot}$ (see Eqs.(\ref{zetatot}) and (\ref{parameter})). 
Then the behavior of $\varphi$ determines the 
evolution of the curvature perturbation which is 
independent of $d_0$. 
The evolution equation for $\varphi_{c}$ is obtained 
from $\delta P_c=c_{s c}^2 \delta \rho_c$ and it is given
by replacing $H,a,c_s^2,d/dt,$ and $\delta \rho_{\E}$
to  $H_c=e^{d_0}(H-\dot{d}_0), a_c=a e^{-d_0}, c_{s c}^2, e^{d_0} d/dt$ and 
$e^{4 d_0} \delta \rho_{\E}$, respectively, in Eq.(\ref{eqphi0}). 
The function $\varphi$ and $\varphi_c$ describe the displacement 
of our brane and the second brane respectively, and their 
relative difference causes the radion perturbation $N$.   
The equation $\Phi+\Psi=-\kappa_4^2 k^{-2} a^2 \delta \pi_{\E}$
gives the evolution equation for $E$;
\begin{eqnarray}
\ddot{E} &+& \left(3H +\frac{2 \dot{d}_0}{e^{2 d_0}-1}\right) \dot{E}
-\frac{1}{3} k^2 a^{-2} E \nonumber\\
&=& -\frac{2 e^{2 d_0}}{e^{2 d_0}-1} k^2 a^{-2} 
(\varphi- e^{-2 d_0} \varphi_c).
\end{eqnarray}
The function $E$ is identified with the bulk anisotropic perturbation 
\cite{Koyama}.  We have a closed set of equations 
for $\varphi, \varphi_c$ and $E$. Therefore, the solution for 
$\delta \pi_{\E}$ can be obtained from Eqs.(\ref{SolpiE}) and 
(\ref{parameter}). 

\hspace{1cm}\\
{\it 5. CMB anisotropy in a simple model}\\
Let us consider the simplest case where $d_0=d_{\ast}=const.$
and $d_{\ast}$ is sufficiently large $e^{-2d_{\ast}} \ll 1 $. 
It can be realized by taking $\rho_c=-\rho e^{2 d_{\ast}}$
and $w=w_c$. This choice is consistent with $\rho_{\E}=0$. 
First, let us consider large scale perturbations. 
Assuming the scale factor is given by $a \propto t^{2/3(1+w)}$
($w=$const.), the solution for $\varphi$ is obtained as 
\begin{eqnarray}
\varphi &=& -\frac{3(1+w)}{3w+1}\zeta_{\ast}-
\frac{1}{9w-1}\left(\frac{\rho_r}{\rho} \right) \delta C_{\ast}, \nonumber\\
\varphi_c &=& -\frac{3(1+w)}{3w+1}\zeta_{c \ast}-
\frac{1}{9w-1}\left(\frac{\rho_r}{\rho} \right) 
e^{2 d_{\ast}}\delta C_{\ast},\nonumber
\end{eqnarray}
for $w \neq -1/3,1/9$, where $\zeta_{c \ast}$ is the curvature perturbation 
on hypersurface of uniform matter energy density on the second brane.
Here we neglected the homogeneous solutions. 
Note that $w=1/9$ is not singular and one can find a solution for $\varphi$
as $-(1/2) \delta C_{\ast} (\rho/\rho_r) \ln a$. 
One point is that $\varphi_c$ that depends on $\delta C_{\ast}$ satisfies
$\varphi_c=e^{2 d_{\ast}} \varphi$, then $\delta C_{\ast}$
does not contribute to $E$. Now we can obtain the solution for
metric perturbations. We find the parts of $\Phi$ and $\Psi$
corresponding to each term in the solution for $\varphi$:
\begin{eqnarray}
\Psi &=& \Psi_{\zeta}+\Psi_{\E}, \quad \Phi=\Phi_{\zeta}+\Phi_{\E},
\quad \Psi_{\E} =  \Phi_{\E}=-\varphi_{\E}, \nonumber\\
\varphi_{\E} &=& -\frac{1}{9w-1}\delta C_{\ast}
\left(\frac{\rho_r}{\rho} \right).
\label{solmetric}
\end{eqnarray}
Here, $\Phi_{\zeta}$ and $\Psi_{\zeta}$ are the same as conventional 
4D solutions except for the additional terms which depend
on the radion $d_{\ast}$ and "shadow matter" $\zeta_{c \ast}$ \cite{Koyama}. 
These additional terms can be neglected for $e^{-2d_{\ast}} \ll 1$. 
Then we find the solution for Weyl anisotropic stress as 
\begin{equation}
\kappa_4^2 k^{-2} a^{2} \delta \pi_{\E} = 2 \varphi_{\E}.
\label{solpi}
\end{equation}

Large scale CMB anisotropy can be written as  
\begin{equation}
\frac{\Delta T}{T}= -\frac{1}{5} \zeta_{\ast}.
\end{equation} 
Dark radiation does not affect $\Delta T/T$.
This is quite a non-trivial result. As mentioned before,
the Weyl energy density perturbations induce the entropy
perturbations. The effect of Weyl anisotropic stress $\delta \pi_{\E}$
exactly cancels this entropy perturbation in Eq.(\ref{SWaniso}). 

Let us investigate whether this cancellation holds for small scale
perturbations.
Under a tight coupling approximation of baryon-photon fluid,
the evolution equation for radiation temperature anisotropy
$\Theta_0$ becomes \cite{HuSugiyama}
\begin{eqnarray}
\ddot{\Theta}_0  &+& H \dot{\Theta}_0+\frac{\dot{R}}{1+R} 
\dot{\Theta}_0+k^2 a^{-2} c_s^2\Theta_0  = F, \nonumber\\
F &=& -\ddot{\Phi}- H\dot{\Phi}-\frac{\dot{R}}{1+R} \dot{\Phi}
-\frac{k^2}{3} a^{-2} \Psi,
\label{eqtemp}
\end{eqnarray}
where $R=3 \rho_b/4 \rho_r$ and $\rho_b$ is the baryon energy density.
The solutions for metric perturbations are given by the upper equation
in Eq.(\ref{solmetric}) though $\varphi_{\E}$ is the solutions 
for Eq.(\ref{eqphi0}) with $E=0$ and the equation for $\delta \rho_{\E}$ 
is obtained by Eqs.(\ref{eqP}) and (\ref{solpi}).
The source term $F$ can be decomposed as $F=F_{\zeta}+F_{\E}$ 
according to the solutions for metric perturbations (Eq.(\ref{solmetric})). 
Observed temperature anisotropy (Eq.(\ref{SW})) is given by 
$\Delta T/T=\Theta_0+\Psi=\Theta_0-\varphi_{\E}+\Psi_{\zeta}$. 
Thus if $\Theta_{0e}\equiv \Theta_0-\varphi_{\E}$
behaves in the same way as conventional 4D theory, the temperature
anisotropy also behaves in the same way. From Eq.(\ref{eqtemp}),
we get the equation for $\Theta_{0e}$:
\begin{eqnarray}
\ddot{\Theta}_{0e} &+& H \dot{\Theta}_{0e}+\frac{\dot{R}}{1+R} 
\dot{\Theta}_{0e}+k^2 a^{-2} c_s^2 \Theta_{0e}  \nonumber\\
&=& F_{\zeta} 
+ \left(\frac{1}{3}-c_s^2 \right)k^2 a^{-2} \varphi_{\E}.
\end{eqnarray}
Hence at large scales $k \to 0$ or in the radiation dominated universe 
$c_s^2=1/3$, the 
temperature anisotropy is exactly the same as the conventional
4D theory. But as the Universe becomes 
matter dominated $c_s^2 \neq 1/3$ and perturbations enter the horizon, 
the behavior of temperature anisotropy is modified. 
Because the amplitude of $\varphi_{\E}$ decreases with time 
under the horizon, the effect is largest for the first 
acoustic oscillation.  
Fig.1 shows the resultant CMB anisotropy for given cosmological 
parameters. One can see that the dark radiation does not modify 
the CMB anisotropy at $l \sim 2$ and at large $l$ but it modifies a 
first acoustic oscillation. 
The resultant spectrum is quite different from 
the prediction of simple mixtures of 
adiabatic and isocurvature perturbations due to Weyl anisotropic 
stress $\delta \pi_{\E}$.

\hspace{1cm}\\
{\it 6. Discussions}\\
In this letter, we developed a formulation to calculate CMB anisotropy
in two branes model at low energies with an 
appropriately small distance between two branes.
In realistic models, we should introduce 
stabilization mechanisms of the radion. 
Our formulation can be extended to include stabilization mechanisms.
We considered only the classical theory of the perturbations. Thus the 
initial spectrum remains to be determined, and it can be modified by 
the effects from the bulk in the brane world. These issues 
need further investigations. 

\begin{figure}[hb]
\centerline{
\includegraphics[width=7.5cm]{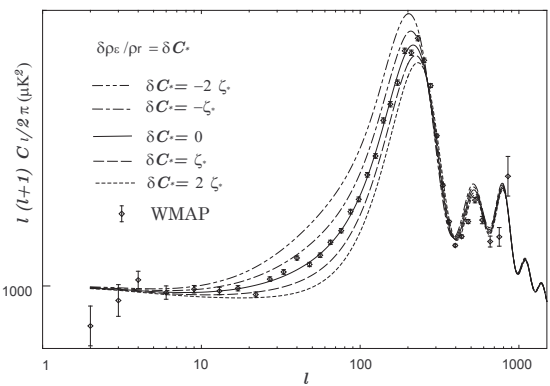}}
\caption{CMB angular power spectrum for various $\delta C_{\ast}$.
$\zeta_{\ast}$ is appropriately normalized. 
We take $\Omega_0=0.26, \Omega_{\Lambda}=0.70, \Omega_b=0.04, h=0.72$ and 
$n=1$. The observational data was taken from WMAP \cite{WMAP} and 
the spectrum was calculated by modifying CMBEASY which is based on 
CMBFAST \cite{code}.}
\label{fig:CMB}
\end{figure}


\begin{references}

\bibitem{RS}
L. Randall and R. Sundrum, Phys. Rev. Lett. {\bf 83}, 3370 (1999);
ibid, {\bf 83}, 4690 (1999). 

\bibitem{WMAP}
C.L. Bennett et. al, astro-ph/0302207.
 
\bibitem{perturbations}
S. Mukohyama, Phys. Rev. D {\bf 62}, 084015 (2000);
H. Kodama, A. Ishibashi and O. Seto, Phys. Rev. D {\bf 62}, 064022 (2000) ;
R. Maartens, Phys. Rev. D {\bf 62}, 084023 (2000) ;
D. Langlois, Phys. Rev. D {\bf 62}, 126012 (2000) ;
C. van de Bruck, M. Dorca, R. H. Brandenberger and 
A. Lukas, Phys. Rev. D {\bf 62}, 123515 (2000) ;
K. Koyama and J. Soda, Phys. Rev. D {\bf 62}, 123502 (2000) .

\bibitem{review}
see, for example, N. Deruelle, Proceedings of the Porto JENAM conference 
(gr-qc/0301035); R. Maartens, Prog. Theor. Phys. Suppl.
{\bf 148}, 213 (2003).

\bibitem{low}
C. Csaki, M. Graesser, L. Randall and J Terning, 
Phys. Rev. {\bf D62}, 045015 (2000);

\bibitem{Koyama}
K. Koyama, Phys. Rev. D {\bf 66} 084003 (2002).

\bibitem{low2}
T. Wiseman, Class. Quantum Grav. {\bf 19}, 3083 (2002);
S. Kanno and J. Soda, Phys. Rev. {\bf D66},083506 (2002).

\bibitem{ShiromizuKoyama}
T. Shiromizu and K. Koyama, Phys. Rev. {\bf D67}, 084022 (2003).

\bibitem{SMS} 
T. Shiromizu, K.I. Maeda and M. Sasaki, Phys. Rev. {\bf D62}, 024012 (2000);

\bibitem{large}
D Langlois, R Maartens, M Sasaki, D Wands,
Phys.Rev. {\bf D63}, 084009 (2001).

\bibitem{BM}
J. D. Barrow and R. Maartens, Phys. Lett. {\bf B532}, 153 (2002).
 
\bibitem{ST}
The CMB anisotopies in scalar-tensor theory have been investigated 
intensively, see for example, R. Nagata, T. Chiba and N. Sugiyama, 
Phys. Rev. {\bf D66}, 103510, (2002); X. Chen and M. Kamionkowski, Phys. Rev. 
{\bf D60}, 104036 (1999). However, the initial conditions for scalar 
field perturbations in these literatures correspond to 
$\delta C_{\ast}=0$ in the brane world context. Thus we cannot apply 
their results for the investigation of the effect of dark radiation perturbation. 
 
\bibitem{HuSugiyama}
W. Hu and N. Sugiyama, Astrophys. J. {\bf 444}, 489 (1995).

\bibitem{code}
M. Doran, astro-ph/0302138; U. Seljak and M. Zaldarriaga, Astrophys. J. 
{\bf 459}, 437 
(1996).

\end{references}
\end{document}